 \documentclass[aps,prl,twocolumn,groupedaddress,showpacs]{revtex4}

\usepackage{graphicx}

\begin{document}



\title{Enhanced Production of Neutron-Rich Rare Isotopes 
       in Peripheral Collisions  \\  at Fermi Energies   } 

\author{G.A. Souliotis}

\author{M. Veselsky}

\author{G. Chubarian}
\author{L. Trache} 
\author{A. Keksis}
\author{E. Martin}
\author{D.V. Shetty}
\author{S.J. Yennello}

\affiliation{Cyclotron Institute, Texas A\&M University, College Station, TX 77843 }

\date{\today}


\begin{abstract}

A large enhancement in the production  of  neutron-rich projectile  
residues  is observed in  the reactions of a 25 MeV/nucleon 
$^{86}$Kr beam with the  neutron rich  $^{124}$Sn and  $^{64}$Ni targets
relative to the predictions of the  EPAX  parametrization of 
high-energy fragmentation, as well as relative to the reaction with the less neutron-rich 
$^{112}$Sn target. The data demonstrate  the significant effect of the target
neutron-to-proton ratio  (N/Z) in peripheral collisions at Fermi energies.
A hybrid model based on a deep-inelastic  transfer code (DIT) followed by a statistical 
de-excitation code  accounts  for part of  the observed large  cross sections.
The DIT simulation  indicates  that the production of  neutron-rich nuclides in these 
reactions is  associated with  peripheral  nucleon exchange 
in which  the  neutron skins of the neutron-rich $^{124}$Sn and $^{64}$Ni 
target nuclei may  play an important  role.
From a practical viewpoint, such  reactions between massive neutron-rich nuclei
offer  a novel   
synthetic avenue to access  extremely neutron-rich 
rare isotopes towards the  neutron-drip line.

\end{abstract}


 \pacs{25.70.-z, 25.70.Hi,25.70.Lm}

 \keywords{ Rare isotope production, neutron-rich nuclides, nuclear reactions, 
            deep inelastic transfer,  Fermi energy, peripheral collisions, neutron skin }

\maketitle


Exploration  of the nuclear landscape towards  the neutron-drip line  
\cite{ndrip1}
is currently of great interest in order to elucidate the evolution of nuclear structure 
with increasing neutron-to-proton ratio (N/Z) \cite{Casten,ngiant} and 
understand important nucleosynthesis pathways \cite{science1}, most notably 
the r-process \cite{rproc}.
Reactions induced by   neutron-rich nuclei provide invaluable information on
the isospin dependence of the nuclear equation of state \cite{science2,Bao1}.  
Extremely  neutron-rich nuclei offer the unprecedented opportunity to extrapolate our 
knowledge to the properties of bulk isospin-rich matter, such as 
neutron stars \cite{nstar1,nstar2}. 
The efficient production of very neutron-rich nuclides
is a key issue in current and future rare isotope beam facilities around the world
\cite{RIA1,RIA2,Z_EURISOL}    
and, in parallel, the search for new synthetic approaches is of exceptional  importance.

Neutron-rich nuclides  have traditionally been produced  in
spallation reactions, fission, 
and projectile  fragmentation \cite{Geissel}.
In high-energy  fragmentation reactions, the production of 
the most neutron-rich isotopes is based on  a ``clean-cut'' removal
of protons from the projectile. The world's data on fragmentation 
cross sections are well represented by the empirical parametrization
EPAX \cite{EPAX}. EPAX  is currently the common basis for predictions 
to plan  rare beam experiments and facilities.   
In addition to the widely used projectile fragmentation approach, 
neutron-rich nuclides can be produced  in multinucleon transfer 
reactions \cite{Corradi}  and
deep-inelastic reactions near  the Coulomb barrier 
(e.g. \cite{Volkov,Broda,IYLee}). In such reactions,
the target N/Z significantly affects  the production
cross sections, but the low velocities of the fragments  and
the  ensuing  wide angular and ionic charge state distributions
render practical applications rather limited.

The  Fermi energy regime (20--40 MeV/nucleon) \cite{Fuchs} offers 
the unique 
opportunity to combine the advantages  of both  low- and high-energy
reactions. At this energy, the synergy of  projectile and  target enhances the N/Z
of the fragments, while the velocities are  high enough to 
allow efficient in-flight collection and separation. 
Early work on neutron-rich fragment production  from heavy 
projectiles  at Fermi energies \cite{Borrel,Bacri,Bazin}
provided  no  production cross sections,  hampering quantitative comparisons
to parametrizations or reaction simulations. 
In order to explore the possibilities offered at  Fermi energies,
we recently undertook an experimental study of the  production cross sections
of projectile residues
from the reaction  25 MeV/nucleon $^{86}$Kr+$^{64}$Ni,  in which  enhanced  production 
of neutron-rich fragments was  observed \cite{George}. 

In this Letter, we focus on  a subsequent detailed study  of near-projectile residues from 
the reactions of $^{86}$Kr (25 MeV/nucleon) with $^{124}$Sn and  $^{112}$Sn targets
and present  a  comparison with the  results from  $^{86}$Kr+$^{64}$Ni \cite{George}, 
as well as with    the expectations of  EPAX \cite{EPAX}. 
We demonstrate that the production of  neutron-rich  projectile  residues is substantially 
enhanced  in  the reactions involving the most neutron-rich systems and we provide 
an interpretation  based on a deep inelastic transfer model.


The 
measurements were performed at the Cyclotron Institute of Texas A\&M
University, following the experimental scheme of our  previous work 
\cite{George}. A concise description is given in the following.
A  25 \hbox{MeV/nucleon} $^{86}$Kr$^{22+}$ beam  ($\sim$1 pnA) from 
the K500 superconducting cyclotron  interacted with  $^{124}$Sn and  $^{112}$Sn
targets (2 mg/cm$^{2}$).
Projectile residues were analyzed with  the MARS recoil separator \cite{MARS}
offering an angular acceptance of 9 msr and momentum acceptance of 4\%.
The beam struck the target at 
4.0$^{o}$ relative to the optical  axis of MARS. Thus, fragments  were accepted in
the polar angular range  2.7$^{o}$--5.4$^{o}$   (lying inside the grazing angle of
6.5$^o$ for the Kr+Sn reaction at 25 MeV/nucleon). 
At the MARS focal plane, the fragments were collected in a  
two-element ($\Delta $E, E)  Si detector telescope. 
Time of flight was measured between two PPACs (parallel plate avalanche 
counters) placed at  the MARS dispersive image and at the focal plane,
respectively.
The horizontal position from the first PPAC  and  the field   
of the  MARS first dipole magnet  determined  the magnetic rigidity,  
$B\rho$. 
The fragments  were  characterized event-by-event by
energy-loss, residual energy, time of flight, and $B\rho$. 
With the procedures described in Ref. \cite{George}, the atomic number Z, 
the ionic charge q, the mass number A  and the velocity   
were obtained with high resolution (0.5, 0.4, 0.6 units and 0.3\%, respectively).
After summation over ionic  charge states
(with corrections for missing charge states), 
fragment cross sections with respect to  Z, A and velocity
were obtained in the angular range 2.7$^{o}$--5.4$^{o}$ and $B\rho$ range
1.3--2.0 T\,m. 
To extract  total  cross sections, the measured yield data 
were corrected  for the limited  angular acceptance
and magnetic rigidity range  covered.  
The  corrections were  obtained from the  ratio of the total 
to the filtered cross sections (with angular and momentum acceptance cuts)
calculated using the simulation approach  described below. 
The corrections involved  an azimuthal angular factor of 10  and  mass-dependent polar factors
ranging from  2 to  4 for  A=85 to 50, respectively (resulting in,  e.g., a total factor of 20
for fragments very  close to the projectile).
The systematic uncertainty  in the extracted cross sections 
is estimated  to be a factor of  2 \cite{George}.


    \begin{figure}[h]                                        

    \includegraphics[width=0.47 \textwidth, height=0.56\textheight ]{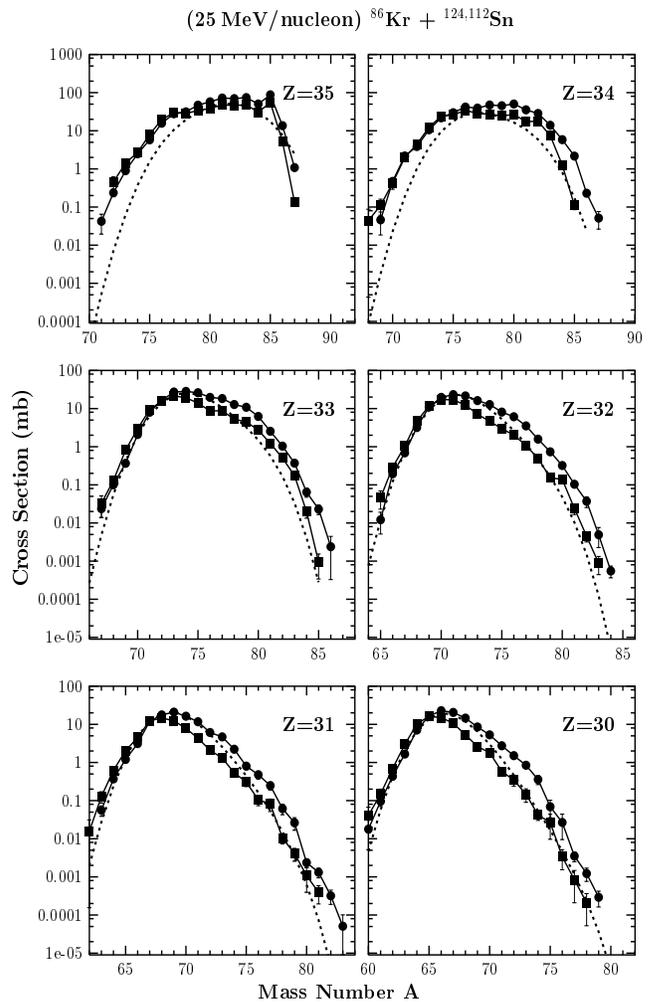}

    \caption{
           Mass distributions of elements Z=30--35 from the reaction of 
           $^{86}$Kr (25MeV/nucleon) with $^{124}$Sn and $^{112}$Sn.
	   The data are shown by full circles for $^{86}$Kr+$^{124}$Sn
           and full squares for $^{86}$Kr+$^{112}$Sn.
           The dotted  lines are   
	   EPAX expectations \cite{EPAX}. 
            }
    \label{ad}
    \end{figure}


    \begin{figure}[h]                                        

    \includegraphics[width=0.40\textwidth, height=0.40 \textheight ]{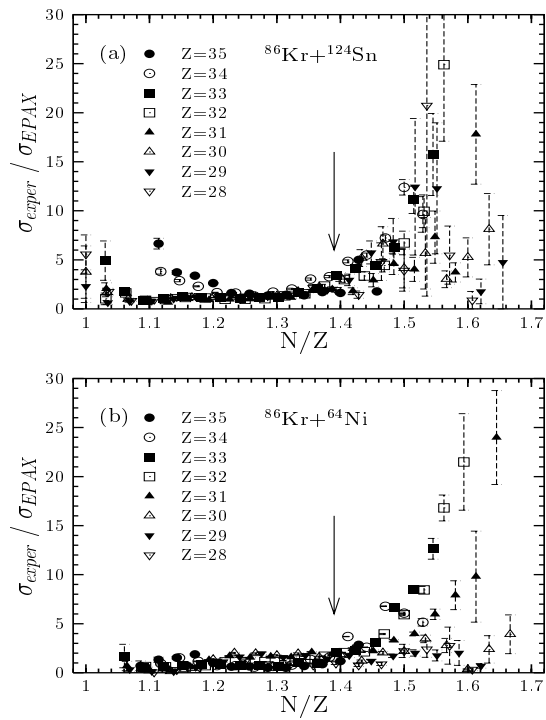}

    \caption{
           Ratio of measured cross sections of projectile residues from 
	   $^{86}$Kr (25MeV/nucleon) on  $^{124}$Sn target (a) (current measurement), 
           and on $^{64}$Ni  target (b) \cite{George}  with respect to
	   EPAX expectations \cite{EPAX}. Arrows indicate the N/Z
           of the projectile. 
            }
    \label{epax}
    \end{figure}


Fig. 1  shows the results of the mass distributions (cross sections) of elements Z=30--35
from the reactions of $^{86}$Kr (25 MeV/nucleon) with $^{124}$Sn (full circles) and
$^{112}$Sn (full squares).
The dotted  lines are  the predictions of the EPAX  parametrization \cite{EPAX}.
We observe that the neutron-rich projectile fragments produced in   the reaction with the more
neutron-rich $^{124}$Sn (N/Z=1.48) target  have considerably larger yields
compared to those obtained in the reaction with the less neutron-rich 
$^{112}$Sn target (N/Z=1.24). Typical ratios  are 3--10 for the most  neutron-rich isotopes observed
in both reactions.  In most of the cases of the present data, the most neutron-rich 
nuclides have been observed only in  the $^{86}$Kr+$^{124}$Sn reaction.
Compared to EPAX, the yields from  $^{86}$Kr+$^{124}$Sn are substantially
larger,  whereas those from  $^{86}$Kr+$^{112}$Sn are in reasonable agreement  with EPAX.
It should be pointed out  that EPAX  represents  an abrasion-ablation type of fragmentation scenario
in which the target is a mere spectator  having (almost) no effect on the production 
cross section  (apart from a geometrical factor) \cite{EPAX}.
To quantify the comparison between the present data and EPAX, we show in Fig. 2a the ratio
of the measured cross sections of  fragments with Z=28--35 from $^{86}$Kr+$^{124}$Sn 
with respect to EPAX expectations.
The same ratio is presented in Fig. 2b for our previous data on  $^{86}$Kr+$^{64}$Ni \cite{George}. 
In the EPAX calculation, we extended the  prediction up to one neutron pick-up products to allow
comparison with the present data. However, our data on  $^{86}$Kr+$^{124}$Sn and
$^{86}$Kr+$^{64}$Ni show the production of nuclides with up to 3--4 neutrons  picked-up from the
 target,  along with usual proton removal products. 
In both figures,  we observe that for nuclei far from the projectile
(Z=28--30), as well as for N/Z less than that of the projectile (N/Z=1.39, arrow 
in Fig. 2), the experimental cross sections are in remarkable agreement with EPAX 
(within a factor of $\sim$2).
However, for heavier elements (Z=31--35) and for progressively higher N/Z, 
a striking increase in the ratio is observed, especially for near-projectile products
involving neutron pick-up.
The observed yield amplification over EPAX  for neutron-rich nuclei
demonstates the dramatic  effect of the neutron-rich target on the production
mechanism at  Fermi energies,  which      
 involves substantial nucleon exchange \cite{Martin}.

To gain  insight into the mechanism of neutron-rich 
nuclide production at this energy,
we performed  simulations  using a  hybrid Monte Carlo approach,
as  in \cite{George}.  The dynamical stage of the collision was  described
by  the deep inelastic
transfer (DIT) code of Tassan-Got \cite{DIT}  and the de-excitation stage was simulated
by the statistical code  GEMINI \cite{GEMINI}.
Comparison of the DIT/GEMINI predictions  with the  data  is shown in Fig. 3 for Se (Z=34),
Ge (Z=32) and Zn (Z=30) (thick solid lines). 
We observe that the calculation describes  well, in most cases, the central and 
the neutron-deficient part of the measured distributions. However, it cannot fully account for 
the  enhancement observed on the neutron-rich sides   
for near-projectile elements.
Backtracing of the DIT simulations indicates  that  neutron-rich products originate
in  peripheral collisions 
in which the projectile--target density distributions overlap by up to 1--1.5 fm.
For larger projectile--target overlaps, the observed enhancement diminishes 
and the  cross sections agree with DIT/GEMINI \cite{George}
and  EPAX (Fig. 2).
It is worth noting  the large cross sections from $^{86}$Kr+$^{64}$Ni \cite{George}
relative to $^{86}$Kr + $^{112}$Sn, although the  targets have nearly  equal N/Z.
In this case it is their position with respect to the valley of  $\beta$-stability
($^{64}$Ni is neutron-rich and   $^{112}$Sn is neutron-poor)
that determines the nucleon flow.
Consequently, the potential energy surface (appropriately  taken into account in the 
DIT code) plays  a definitive  role in  the production of   neutron-rich
fragments at Fermi energies.


    \begin{figure}[h]                                        

    \includegraphics[width=0.47\textwidth, height=0.50\textheight ]{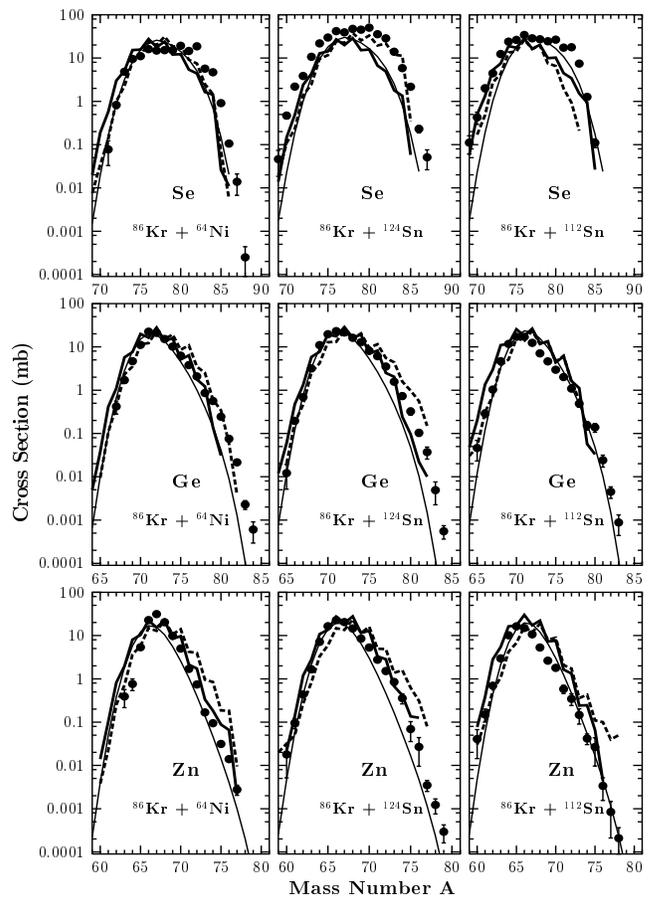}

    \caption{
           Comparison  of experimental mass distributions of Se (Z=34),
           Ge (Z=32) and Zn (Z=30) from $^{86}$Kr (25 MeV/nucleon) on 
           $^{64}$Ni \cite{George}, $^{124}$Sn and $^{112}$Sn (present data)
           with DIT/GEMINI calculations without (thick solid line), and
           with (thick dashed line) neutron and proton
           density distributions.
           The thin solid line is  the EPAX prediction \cite{EPAX}.
           }
    \label{calc_skin}
    \end{figure}


We should point out that, in the original DIT code \cite{DIT}, the  nuclei are assumed 
spherical with homogeneous neutron and proton  density distributions.
However, it is well established  that neutron-rich nuclei possess neutron-skins; namely,
their neutron density distributions extend further out relative to the proton distributions.
As well known, the neutron skin is predicted by a variety of theoretical models
including microscopic Thomas-Fermi, Skyrme Hartree-Fock and relativistic
mean-field approaches (e.g. \cite{Andreas},\cite{Brown},\cite{RMF}). 
Also, it has recently  been demonstrated experimentally in annihilation studies 
of antiprotonic atoms \cite{antiproton}.  
Some time ago, Harvey \cite{Harvey} took into consideration 
the neutron skin of heavy targets ($^{208}$Pb, $^{197}$Au) 
in  a microscopic  nucleon-nucleon scattering model  to 
explain the enhanced production of neutron-rich projectile  fragments
from  $^{16}$O and $^{20}$Ne at Fermi energies relative to  those
at relativistic energies. His approach also served as  an  
interpretation of the N/Z properties of projectile fragment yield data from
$^{40}$Ar (44 MeV/nucleon) on heavy targets  \cite{Borrel}.
Although the DIT  code employed in our simulations   is based on  nucleon exchange
(rather than  nucleon-nucleon collisions)   in the overlap  zone,
we may  attribute its partial success to describe the observed enhanced cross sections
to the inadequate  treatment of peripheral nucleon exchange,   in which  the neutron skins
of the reaction partners may  play important roles.
To investigate this assumption further, we  included
neutron and proton density distributions  into the  DIT code.
The density distributions were calculated with a  microscopic Thomas-Fermi
code \cite{Andreas} and  employed  in  DIT.  In the  simulation, 
at a given projectile-target overlap
and a given  nucleon exchange site,
the transfer probabilities were scaled by the local peripheral 
(``halo'')  factors, $ (\rho_{n}/\rho_{p}) / (N/Z) $, of the respective
reaction partners. The change of the density profiles was followed during 
the (sequential) nucleon exchange.
The results of this  modified-DIT/GEMINI calculation are shown in Fig. 3  by
dashed lines.
A qualitative improvement in the description of the  neutron-rich sides 
of the distributions for near-projectile elements from the very neutron-rich systems   
$^{86}$Kr+$^{64}$Ni  and $^{86}$Kr+$^{124}$Sn is found.     
However, for the  $^{86}$Kr+$^{112}$Sn reaction, the standard DIT/GEMINI calculation
agrees   better with the data.
These calculations indicate, although qualitatively, the possible  effect of the  
neutron skin   of the neutron-rich targets
$^{64}$Ni and $^{124}$Sn in the observed enhancement.
Guided by these observations,  we  infer  that an improved  deep-inelastic
transfer model, or an equivalent approach, which self-consistently  takes into account 
the neutron and proton density distributions of the reaction partners 
may  account for the  large cross sections of the neutron-rich
products observed in such reactions.

From a practical viewpoint,   the large production  cross section
of neutron-rich fragments  indicate  that  such  reactions  offer an attractive
approach to produce very  neutron-rich  nuclides.
Apart from direct in-flight or IGISOL-type options, we propose  the application
of such  reactions  as a second stage in two-stage rare beam production schemes.
For example, a beam of $^{92}$Kr  from an ISOL-type facility (or, from the proposed 
Rare Isotope Accelerator Facility (RIA) \cite{RIA1,RIA2}) can be 
accelerated around the  Fermi energy and interact with a very  neutron-rich target
(e.g. $^{64}$Ni, $^{124}$Sn, $^{208}$Pb, $^{238}$U) to produce 
extremely  neutron-rich nuclides
that cannot be accessed by fission or
projectile fragmentation.
A quantitative prediction of rates of such nuclides  will be possible after
further experimental studies and  improvement of the  description of peripheral 
collisions  between very neutron-rich heavy nuclei in the Fermi energy regime.
   

In summary, a substantial enhancement in the production cross sections of 
neutron-rich projectile   fragments is observed in  the reactions 
$^{86}$Kr (25 MeV/nucleon)+$^{124}$Sn and
$^{86}$Kr (25 MeV/nucleon)+$^{64}$Ni
relative to the EPAX  parametrization, as well as 
relative to the less neutron-rich system  $^{86}$Kr+$^{112}$Sn.
The  hibrid DIT/GEMINI model
provides a partial interpretation of the observed cross sections.
The DIT simulations indicate  that the enhanced production of  neutron-rich isotopes
in the reactions $^{86}$Kr+$^{124}$Sn and  $^{86}$Kr+$^{64}$Ni
is  associated with  peripheral nucleon exchange. In such collisions,
the  neutron skins  of the  $^{124}$Sn and $^{64}$Ni target nuclei
may   play   a significant  role. 
For  practical purposes, such reactions
offer a novel and competitive  pathway  to  produce extremely neutron-rich 
isotopes towards the  neutron-drip line.

We are thankful to L. Tassan-Got for his DIT code and useful suggestions. 
We also wish to thank A. Sanzhur for insightful discussions  and R. Charity
for using the GEMINI code. We gratefully  acknowledge the support of the  operations
staff of the Cyclotron Institute during the measurements.
Financial support for this work was provided, in part, by the U.S. Department 
of Energy under Grant No. DE-FG03-93ER40773 and by the Robert A. Welch 
Foundation under Grant No. A-1266.


\bibliography{nskin.bib}



\end{document}